\documentclass[final,5p,times,twocolumn]{elsarticle}

\usepackage{amssymb}

\usepackage{graphicx}  
\usepackage{dcolumn}   
\usepackage{bm}        
\usepackage{amsmath}   

\journal{Phys. Lett. B}

\begin{document}

\begin{frontmatter}




\title{A Novel Solution to the Klein-Gordon Equation in the Presence of a Strong Rotating Electric Field }

\author[huji,soreq]{E. Raicher}
\ead{erez.raicher@mail.huji.ac.il}
\author[soreq,madrid]{S. Eliezer}
\author[huji]{A. Zigler}
\address[huji]{Racah Institute of Physics, Hebrew University, Jerusalem 91904, Israel}
\address[soreq]{Department of Applied Physics, Soreq Nuclear Research Center, Yavne 81800, Israel }
\address[madrid]{Nuclear Fusion Institute, Polytechnic University of Madrid, Madrid, Spain }

\begin{abstract}
The Klein-Gordon equation in the presence of a strong electric field, taking the form of the Mathieu equation, is studied. A novel analytical solution is derived for particles whose asymptotic energy is much lower or much higher than the electromagnetic field amplitude. 
The condition for which the new solution recovers the familiar Volkov wavefunction naturally follows. When not satisfied, significant deviation from the Volkov wavefunction is demonstrated. The new condition is shown to differ by orders of magnitudes from the commonly used one. 
As this equation describes (neglecting spin effects) the emission processes and the particle motion in Quantum Electrodynamics (QED) cascades, our results suggest that the standard theoretical approach towards this phenomenon should be revised. 
\end{abstract}

\end{frontmatter}


\section{Introduction}
At present days, several laser infrastructures with expected intensity of $10^{24}-10^{25}W/cm^2$ are under construction worldwide\cite{ELI, XCELS, HIPER, GEKKO}.
The experimental availability of such intense field sources creates exciting opportunities in many research fields \cite{Tajima}, such as QED in the presence of strong fields \cite{DiPiazza}, Schwinger mechanism \cite{schwinger, narozh, popov, heinzl, heben}, Unruh radiation \cite{unruh1, unruh2}, novel fast ignition schemes \cite{Tajima,FI, FI2, shalom}, particles acceleration \cite{ion acc}, high harmonics generation \cite{atto} as well as nuclear physics and the search for dark matter candidates \cite{mendonca2, chavez}.

The fundamental physics underlying all these scientific applications is the interaction of an intense electromagnetic (from now on we shall use the initials EM) field with an electron.
The nature of the interaction is determined by the normalized field amplitude $\xi \equiv ea/m$ and the quantum parameter 
\begin{equation}
\chi \equiv \frac{e}{m^3} \sqrt{-(F^{\mu \nu} p_{\nu})^2}.
\label{eq:av chi}
\end{equation}
 The electron mass and charge are denoted by $m,e$ respectively, and its asymptotic momentum (i.e. the momentum in the absence of the EM field) is $p_{\mu}=(p_0,p_1,p_2,p_3)$. The amplitude of the vector potential $A_{\mu}$ is $a=\sqrt{-A^2}$ and the EM field tensor is $F^{\mu \nu} \equiv \partial^{\mu}A^{\nu} - \partial^{\nu}A^{\mu}$. Natural units are used, namely $\hbar=c=1$, where $\hbar$ is the reduced Planck constant and $c$ is the speed of light.
If both $\xi$ and $\chi$ are larger than one, the electron dynamics is both quantum and non-linear. The appropriate framework is the strong-Field QED \cite{furry}. Its basic principle is the inclusion of the term corresponding to the interaction with the classical laser field into the free part of the Lagrangian. As a consequence, the unpertubed states appearing in the cross section calculation are no longer free waves. Instead, they are represented by the solutions of the quantum equation of motion in the presence of the EM field. These wavefunctions were obtained by Volkov \cite{Volkov}, provided that the EM field propagates in vacuum.

Employing this approach, the properties of QED in the nonperturbative regime were thoroughly investigated through the years \cite{ritus, QED1, QED2, QED3, QED4, QED5, QED6, nousch, muller1, muller2, tuchin}.
The lowest order strong-field processes are the non linear Compton scattering, where an electron interacts with the laser photons and emits a hard photon and the non-linear Breit-Wheeler process, where a photon decayes into a positron-electron pair in the presence of the EM field \cite{ritus}. A sequential series of these processes, called "QED cascade" is followed by a rapid formation of a QED plasma whose ingredients are electrons, positrons and gamma photons \cite{sokolov_prl}. 
Besides its fundamental significance, this phenomena attracts scientific attention as a possible laboratory astrophysics settings \cite{liang} and as a potential gamma ray source \cite{nerush}. Furthermore, it was suggested that spontaneous cascades impose a limit (of about $10^{25}W/cm^2$) on the achievable laser intensity \cite{fedotov}. 

\section{The physical scenario}
The most favorable EM field configuration for the generation of a QED cascade is a rotating electric field \cite{elkina}. It can be realized experimentally in the anti-nodes of counter propagating laser beams.
The kinetic calculation of the cascade formation is based on a Monte Carlo technique describing the quantum processes mentioned above, integrated with a Particle-In-Cell code taking into account the collective EM field influence on the classical motion of the electrons \cite{elkina,sokolov, ridgers, ridgers2, brady, pukhov, nerush2}. The quantum rates were obtained with the Volkov wavefunctions \cite{ritus}, though not formally adequate for a rotating electric field. 

The underlying assumption (see p. 3 in \cite{ritus} and p. 454 in \cite{landau}) justifying this technique is that as long as the normalized EM field invariants, $\mathcal{F} \equiv e^2F_{\mu \nu}F^{\mu \nu}/(4m^4)$ and $ \mathcal{G} \equiv e^2\epsilon_{\alpha \beta \mu \nu}F^{\alpha \beta}F^{\mu \nu} / (4m^4)$, are negligible with respect to $1$ and $\chi^2$, namely
\begin{equation}
\mathcal{F},\mathcal{G} \ll 1, \chi^2
\label{eq:av assump}
\end{equation}
the particle dynamic is well described by the Volkov wavefunction. The symbol $\epsilon_{\alpha \beta \mu \nu}$ above stands for the Levi-Civita tensor.
The aim of this publication is twofold. First, we explicitly demonstrate that this long-believed assumption is incorrect. For this purpose we derive a novel solution to the quantum equation of motion in the presence of a rotating electric field. The comparison between the Volkov wavefunction and our new solution in the relevant parameters range reveals an overwhelming deviation. Consequently, the common theoretical treatment towards the QED cascades has to be revised. Second, the new solution derived below suggests a natural way to obtain the emission rates in the presence of a rotating electric field for the sake of QED cascades calculations.

\section{The Governing Equation}
The particles of interest in the context of QED cascades are fermions (electrons and positrons). Therefore, their wavefunction obeys the Dirac equation. However, the spin effect may be neglected in the typical cascades conditions \cite{elotzky}. Hence, for the sake of simplicity we shall treat them as scalars.
Thus, the free equation of motion of the particle wavefunction $\Phi$ is the familiar Klein-Gordon equation in the presence of a EM field.
\begin{equation}
\left[-\partial^2-2ie(A\cdot\partial)+e^2A^2-m^2 \right]\Phi = 0.
\label{eq:av klein}
\end{equation}
The center dot stands for Lorentz contraction.
The EM field depends upon the spatial and temporal coordinates through $\phi \equiv k \cdot x$ where $k_{\mu}=(\omega_L,0,0,k_z)$ is the wave vector. We assume a circularly polarized field
\begin{equation}
A(\phi) = a(\phi) \left ( \epsilon e^{i\phi} +  {\epsilon}^* e^{-i\phi} \right )
\label{eq:av ve_pot}
\end{equation}
where the polarization vectors are $\epsilon =( e_{1}-ie_{2} ) / \sqrt{2}$ and $^*$ denotes complex conjugate.	 $a(\phi)$ is a slowly-varying amplitude vanishing at $\phi \rightarrow \pm \infty$. However, in the following this envelope is assumed to be slow enough so that $a^2$ is approximately constant.

The most general dispersion relation of the EM field is massive-like
$k^2 = \omega_L^2-\vec{\textbf{k}}^2 \equiv m_{ph}^2$, where $m_{ph}$ is the effective mass of the EM wave photons.
In the wave frame of reference, $k_z = 0$ and the particle experiences a rotating electric field with frequency $m_{ph}$. On the other hand, it is well known that a standing wave formed by two counter-propagating beams takes the form $A \propto \cos(k_z z) \cos(\omega t)$.
Consequently, a particle located in the vicinity of the anti-nodes, where $\cos k_z z \approx 1+ O(k_z^2 z^2)$, also experiences a rotating electric field. As a result, Eqs. (\ref{eq:av klein}, \ref{eq:av ve_pot}) enable us to study the dynamics of an electron placed in the anti-nodes of the standing laser wave, where the wave frame of the former coincides with the lab frame of the latter. It is justified to restrict our discussion to the anti-nodes since in the case of ultra-relativisitc intensity $\xi \gg 1$, the quantum emission processes take place on a length scale much shorter than the laser wavelength \cite{ritus}, meaning that the spatial dependence may indeed be ignored.
Notice that this dispersion also describes an EM wave propagating through plasma, with $m_{ph}$ analogous to the plasma frequency of a classical plasma wave \cite{polovin}.  A rigorous Lagrangian formulation of this plasma wave may be found elsewhere \cite{myPaper2}. 

The wavefucntion is characterized by the asympototic momentum $p$, namely
$\Phi_p = e^{-ipx}F(\phi)$. The substitution of this ansatz into (\ref{eq:av klein}) yields
\begin{multline}
{m_{ph}}^2F'' - 2ie(k\cdot p)F' +\\ \left[e^2a^2 + 2ea (p \cdot \epsilon) e^{i\phi} + 2ea (p \cdot \epsilon)^* e^{-i\phi} \right]F=0.
\label{eq:av G_eq}
\end{multline}
Equation (\ref{eq:av G_eq}) was considered by several authors. In the following the main approaches are briefly reviewed. 
If $(p \cdot \epsilon)$ vanishes the quantum equation of motion may be solved analytically \cite{myPaper1} even with the spin term included (i.e. Dirac equation). A solution for a discrete set of momentum values was derived in \cite{varro1,varro2}. If the EM wave dispersion relation is assumed to be vacuum-like, i.e. $m_{ph}$ vanishes, then (\ref{eq:av G_eq}) reduces to a $1^{st}$ order equation and therefore admits an exact solution - the familiar Volkov wavefunction \cite{Volkov,landau}. As a result, a perturbative expansion with resepect to $m_{ph}$ may yield an approximate solution. Since the small parameter multiplies the highest derivative, singular perturbation techniques are required \cite{bender}. The most appropriate for our problem is the familiar WKB, which was applied by \cite{piazza_sol,mendonca}. 

A different approach was taken by \cite{becker, noga}, showing that (\ref{eq:av G_eq}) is equivalent to the familiar Mathieu equation. 
With the aid of the transformation $F=y(z) \exp \left[i\frac{k \cdot p}{m_{ph}^2} \phi \right]$, one can prove the equivalence of (\ref{eq:av G_eq}) to the Mathieu equation \cite{becker}
\begin{equation}
y'' +  \left[\lambda - 2q \cos 2z\right]y=0
\label{eq:av mathieu}
\end{equation}
where the following relations are used
\begin{equation}
\lambda \equiv \frac{4}{m_{ph}^2} \left [ \frac{\left( k \cdot p \right)^2}{m_{ph}^2} +(ea)^2 \right], \quad  q \equiv -\frac{8}{m_{ph}^2} ea |p \cdot \epsilon|
\label{eq:av lambda}
\end{equation}
and 
\begin{equation}
z \equiv [\phi + \phi_0]/2, \quad p \cdot \epsilon = |p \cdot \epsilon|e^{i \phi_0}.
\label{eq:av Z_def}
\end{equation}
The tag symbol stands for derivative with respect to $z$.

In this paper, we construct an approximate solution to (\ref{eq:av mathieu}) employing a novel mathematical technique. We shall prove that under the condition $q / \lambda \ll 1$ it is equivalent to an effective $1^{st}$ order equation and thus is easy to solve.

\section{The Novel Solution}
We start by reviewing the standard way to solve numerically the Mathieu equation \cite{mclahan,stegun}. This formalism will prove useful for the sake of our novel derivation.
Eq. (\ref{eq:av mathieu}) bears an apparent similiarity with the equation of motion of an electron in a crystal, if $A_{\mu}(\phi)$ is replaced by the periodic potential.
Hence, the Floquet theory (analogous to Bloch theorem in solid state) is applicable.
Accordingly, the solution may be expressed as $y = P(z)e^{i\mu z}$ where $P(z)$ is a periodic function and $\mu$ is called the characteristic exponential.
Due to its periodicity $P(z)$ may be expanded in a Fourier series.
\begin{equation}
y(z) = e^{i\mu z} \sum_n{c_{2n}}e^{2izn}
\label{eq:av series}
\end{equation}
where $c_{2n}$ are the spectral coefficients.
Substituting (\ref{eq:av series}) into (\ref{eq:av mathieu}) yields
\begin{equation}
V_{2n}c_{2n} = c_{2n+2}+c_{2n-2}
\label{eq:av recursion}
\end{equation}
where the notation of \cite{stegun} is adopted, i.e.
\begin{equation}
V_{2n} \equiv \frac{\lambda - (2n+\mu)^2}{q}.
\label{eq:av Vn}
\end{equation}
Dividing (\ref{eq:av recursion}) by $c_{2n-2}$ we have
\begin{equation}
V_{2n}G_{2n} = G_{2n}G_{2n+2}+1
\end{equation}
where $G_{2n} \equiv c_{2n}/c_{2n-2}$.
Hence, $G_{2n}$ is given by 
\begin{equation}
G_{2n} = \frac{1}{V_{2n} - G_{2n+2}}.
\label{eq:av recursion2}
\end{equation}
It can be shown that for sufficiently large $n$, the dominant term in the denominator of (\ref{eq:av recursion2}) is $V_{2n}$.
Since $V_{2n}$ is growing polynomially in $n$, it follows that $c_{2n}$ decays rapidly in $n$. Consequently, the series (\ref{eq:av series}) may be trancated at a certain index denoted by $n^*$.
Hence, $G_{2n^*}=1/V_{2n^*}$, and the lower harmonics are obtained through the recursion relation (\ref{eq:av recursion2}).
Similiarly, for $n<0$ we have
\begin{equation}
H_{-2n} = \frac{1}{V_{-2n-2} - H_{-2n-2}}
\label{eq:av recursion3}
\end{equation}
where $H_{-2n} \equiv c_{-2n-2}/ c_{-2n}$.
The solution consistency requires the ratios $H_0,G_0$ to be related by
\begin{equation}
H_0 G_0 = 1.
\label{eq:av cond2}
\end{equation}
Iterating this condition, the characteristic exponential $\mu$ is numerically obtained.

This is the starting point of our derivation.
At the moment, the wavefunction spectral width $n^*$ is unknown. Let us assume that it obeys 
\begin{equation}
2n^* \ll \mu.
\label{eq:av delta1}
\end{equation}
Therefore, the quadratic term in $n^2$ appearing in (\ref{eq:av Vn}) is negligible and $V_{2n}$ is anti-symmetric with respect to $n$. As a result, one can verify that $\mu=\pm \sqrt{\lambda}$ yields $V_0 = 0$ and satisfies (\ref{eq:av cond2}). In the following we shall obtain an explicit formula for $n^*$ and thus determine the validity range of our approximation. 
Substituting $\mu=-\sqrt{\lambda}$ in the expression for $V_{2n}$ leads to
\begin{equation}
V_{2n} \approx \frac{4n\sqrt{\lambda}}{q}.
\label{eq:av coef3}
\end{equation}
It should be mentioned that we do not consider the solution that corresponds to the positive root $\nu = \sqrt{\lambda}$ for physical reasons, as shall be discussed below. 

Now let us examine the following $1^{st}$ order equation.
\begin{equation}
y' +i \left[ \sqrt{\lambda} -  \frac{q}{\sqrt{\lambda}} \cos \left(2z \right) \right] y =0.
\label{eq:av effective}
\end{equation}
One can verify that substituting the Floquet ansatz (\ref{eq:av series}) into (\ref{eq:av effective}) yields the recursion relation (\ref{eq:av recursion}) with $\mu = -\sqrt{\lambda}$ and $V_{2n}$ identical to (\ref{eq:av coef3}).
It implies that as long as (\ref{eq:av delta1}) holds, the original $2^{nd}$ order equation is equivalent to an effective $1^{st}$ order one.
The effective equation (\ref{eq:av effective}) admits the straightforward solution
\begin{equation}
y=\exp \left[ -i\sqrt{\lambda} z + i\frac{q}{2\sqrt{\lambda}} \sin (2z)   \right].
\label{eq:av final3}
\end{equation}

The analytical form of (\ref{eq:av final3}) provides us with an expression for the coefficients. Employing the identity \cite{stegun}
\begin{equation}
e^{i X \sin \phi} = \sum J_n(X) e^{in \phi}
\label{eq:av ident}
\end{equation}
where $J_n$ is the Bessel function, we deduce that 
\begin{equation}
c_{2n} = J_n \left(   X   \right), \quad X \equiv \frac{q}{2\sqrt{\lambda}}.
\end{equation}
The Bessel function vanishes for $X=0$ and reaches its first maximum, for $n \gg 1$, at $X \approx n +O(n^{1/3})$ \cite{stegun}. The rise to the peak is extremely rapid, so that $J_n(X)$ is practically zero for $X<n$. In order to establish this statement, we take advatage of the expansion of $J_n(X)$ for $X \le n$ (p. 250 in \cite{watson}, leading term only).
\begin{multline}
J_n(X) \approx \frac{\tanh \alpha (X)}{\pi \sqrt{3}} K_{1/3} \left( \frac{1}{3} n \tanh^3 \alpha(X)  \right) \exp \bigl[ n \bigl( \tanh \alpha(X) + \\ \frac{1}{3} \tanh^3 \alpha(X) - \alpha(X)  \bigr )   \bigr ] 
\label{eq:av nicholson}
\end{multline}
where $K_{1/3}$ is the modified Bessel function and a new quantity is introduced
\begin{equation}
\tanh \alpha \equiv \sqrt{1 - \left( \frac{X}{n} \right)^2 }.
\end{equation}
We are interested with $X \approx n$, namely $\alpha \ll 1$. Hence, we Taylor expand the exponent argument 
\begin{equation}
\tanh \alpha + \frac{1}{3} \tanh^3 \alpha - \alpha \approx -\frac{\alpha^5}{5}
\label{eq:av taylor}
\end{equation}
and employ the asymptotic expression of $K_{1/3}$ for a small argument $u$
\begin{equation}
K_{1/3}(u) \approx \mathcal{C}_1 u^{-1/3}
\label{eq:av mod_bess}
\end{equation}
where $\mathcal{C}_1$ is an insignificant constant. Substituting (\ref{eq:av mod_bess}) and (\ref{eq:av taylor}) into (\ref{eq:av nicholson}) we get
\begin{equation}
J_n(X) \approx \mathcal{C}_2  n^{-1/3}\exp \left[- \frac{n}{5} \alpha^5(X)    \right]
\end{equation}
where $\mathcal{C}_2$ is another constant. Let us write $J_n$ in terms of $\Delta X \equiv n - X$ instead of $\alpha$.
\begin{equation}
\alpha \approx \tanh \alpha \approx \sqrt{1 - \left( 1- \frac{\Delta X}{n} \right)^2 } \approx \sqrt{\frac{2 \Delta X}{n}}.
\end{equation}
Finally, we have
\begin{equation}
J_n(n-\Delta X) \approx \mathcal{C}_2 n^{-1/3}\exp \left[- \frac{n}{5} \left( \frac{2 \Delta X}{n}  \right)^{5/2}    \right].
\end{equation}
That is to say, the decay is exponential and the width is negligible, namely $\Delta X / n \propto n^{-2/5}$.
As a result, ploting the coefficients $c_{2n }(X)$ as a function of $n$ for a given $X$ one observes a rapid decay for $n >X$, as demonstrated in the results section below. Consequently,  the spectral width of the solution (\ref{eq:av final3}) is
\begin{equation}
2n^* \approx \frac{|q|}{\sqrt{\lambda}  }.
\label{eq:av n_star}
\end{equation}
Comparing it with (\ref{eq:av delta1}) we obtain the validity condition, which takes the following form 
\begin{equation}
\delta \equiv 2n^* / \mu \approx \left|\frac{q}{\lambda} \right| \ll 1.
\label{eq:av cond1}
\end{equation}

The exact solution of (\ref{eq:av mathieu}) is known to exhibit instabilities in certain domains of the $(q,\lambda)$ space. Moreover, (\ref{eq:av Vn}) implies that the spectral distribution is essentialy non symmetric due to the term proportional to $n^2$. However, the expression (\ref{eq:av final3}), being a solution of a $1^{st}$ order equation, is spectrally symmetric and always stable. Subsequently, we deduce that as long as (\ref{eq:av cond1}) holds the solution lies in the stable region and the spectral non-symmetry is negligible. 
	
\section{Physical Interpretation}
Let us write the final solution to (\ref{eq:av klein}) with the aid of (\ref{eq:av final3}). The wavefunction $\Phi_p$ is related to $y(z)$ by
\begin{equation}
\Phi_p(x) =y(z) \exp \left[ -i p \cdot x + \phi \frac{(k \cdot p)}{m_{ph}^2} \right].
\end{equation}
We define
\begin{equation}
\nu \equiv \frac{\mu}{2} +\frac{(k \cdot p)}{m_{ph}^2}
\label{eq:av nu_def}
\end{equation}
and use the expressions (\ref{eq:av lambda}, \ref{eq:av Z_def}) relating $q,\lambda$ and $z$ to the physical quantities of the problem.
\begin{equation}
\Phi_p(x) =  \exp \left[ -i \left(p+\nu k \right) \cdot x -i n^* \sin \left( k \cdot x +\phi_0 \right)  \right].
\label{eq:av sol_final2}
\end{equation}
The phase $\phi_0$ is defined in (\ref{eq:av Z_def}) and $n^*, \nu$ read
\begin{equation}
n^* =  \frac{ea|p \cdot \epsilon|}{(k \cdot p) \Omega}, \quad  \nu = \frac{ k \cdot p  }{ m_{ph}^2} \left ( \Omega  - 1   \right )
\label{eq:av n_nu}
\end{equation}
where a new quantity is introduced
\begin{equation}
\Omega \equiv  \sqrt{1 +  \left( \frac{eam_{ph}}{k \cdot p } \right)^2  }  .
\end{equation}
In order to clarify the physical meaning of $n^*,\nu$ we eploit again the Fluquet representation of the wavefucntion
\begin{equation}
\Phi_p(x) =  e^{ -i \left(p+\nu k \right) \cdot x } \sum_n{c_{2n}}e^{in (k \cdot x)}
\label{eq:av sol_final4}
\end{equation}
where the coefficients, in accordance with (\ref{eq:av ident}), take the form
\begin{equation}
c_{2n} = J_n \left(  n^*   \right).
\label{eq:av coeff1}
\end{equation}
The wavefunction (\ref{eq:av sol_final4}) is a superposition of free waves with momenta $p_{\mu}+(\nu+n)k_{\mu}$ weighted according to $c_{2n}$. Each wave may be regarded as an electron carrying $\nu + n$ laser photons.
The quantity $n^*$, as was shown in the previous section, corresponds to the width of the spectral distribution, and $\nu$ may be regarded as its center. In other words, it is the average number of laser photons carried by the electron, leading to the definition of the quasi momentum
\begin{equation}
Q_{\mu} \equiv p_{\mu} + k_{\mu} \nu.
\end{equation}
The effective mass is associated with the quasi-momentum through 
\begin{equation}
m_* \equiv \sqrt{Q^2} = m \sqrt{1+\left(   \frac{ea}{m} \right)^2}.
\end{equation}
Interestingly, the effective mass is the same as in the Volkov case, though $\nu$ is different (as discussed below). $\Omega$ determines the deviation from Volkov. This statement becomes apparent if we recall \cite{landau} that that the Volkov wavefunction takes the same form as (\ref{eq:av sol_final2}) but with 
\begin{equation}
n^*_V = \frac{ea|p \cdot \epsilon|}{(k \cdot p)}, \quad \nu_V = \frac{e^2a^2}{2(k \cdot p)}
\label{eq:av n_nu_V}
\end{equation}
leading to the coefficients
\begin{equation}
c^V_{2n} = J_n \left(  n^*_V   \right).
\label{eq:av coeff2}
\end{equation}
Dividing (\ref{eq:av n_nu}) by (\ref{eq:av n_nu_V}) the following relations are obtained
\begin{equation}
n^* = \frac{n^*_V}{\Omega }, \quad \nu=\frac{2}{\Omega + 1} \nu_V.
\label{eq:av n_nu_comp}
\end{equation}
Taking the limit $m_{ph} \rightarrow 0$ yields $\Omega \rightarrow 1$ so that the Volkov wavefunction is recovered. Now the neglection of the second solution corresponding to $\mu=\sqrt{\lambda}$ may be comprehended. Should we take the positive root, the brackets in (\ref{eq:av n_nu}) would become $(\Omega+1)$. As a result, the limit $m_{ph} \rightarrow 0$ leads to unphysical divergence. In the opposite limit, $\Omega \gg 1$, we have $\Omega \approx eam_{ph} / (k \cdot p)$ and therefore
\begin{equation}
\nu \approx   \frac{ ea  }{ m_{ph}}, \quad n^* \approx \frac{|p \cdot \epsilon|}{m_{ph}}.
\end{equation}
Notice that in this case, as opposed to the Volkov case, $n^*$ is independent of the EM field amplitude and $\nu$ is independent of the particle momentum.

We have mentioned in the introduction that according to the common knowledge the condition (\ref{eq:av assump}) allows one to approximate the wavefunction by the Volkov solution even for a non-vanishing $m_{ph}$.
Let us write down $\chi, \mathcal{F}$ explicitly. For this purpose the vector potential (\ref{eq:av ve_pot}) is substituted into (\ref{eq:av chi}, \ref{eq:av assump}).
\begin{equation}
\mathcal{F} = \left(  \frac{eam_{ph}}{m^2} \right)^2, \quad \chi^2 = \frac{(ea)^2 (k \cdot p)^2 + (eam_{ph})^2 |p \cdot \epsilon|^2}{m^6}.
\label{eq:av chi2}
\end{equation}
Since the second term in the expression for $\chi^2$ is allways smaller than the first, and as we are interested in order of magnitudes only, we have $\chi^2 \approx \frac{(ea)^2 (k \cdot p)^2}{m^6}$. Therefore, the conditions (\ref{eq:av assump}) become
\begin{equation}
\left(  \frac{eam_{ph}}{m^2} \right)^2, \left(  \frac{mm_{ph}}{k \cdot p} \right)^2 \ll1.
\label{eq:av assump3}
\end{equation}
However, Eq. (\ref{eq:av n_nu_comp}) explicitly proves that $m_{ph}$ may be neglected only if $\Omega \approx 1$, or equivalently
\begin{equation}
\Omega^2 - 1 = \left( \frac{eam_{ph}}{k \cdot p } \right)^2 \ll 1.
\label{eq:av assump4}
\end{equation}
Let us plug in typical numbers for a standing wave created by optical laser beams with intensity $I \approx 10^{24} W/cm^2$ (as expected in ELI \cite{ELI}), namely $ea/m = 10^3$ as well as $m_{ph} / m = 10^{-6}$. It is favorable, for this purpose, to evaluate (\ref{eq:av assump3}, \ref{eq:av assump4}) in the wave framework, where $k \cdot p = p_0 m_{ph}$. We obtain
\begin{equation}
\mathcal{F} = \left(  \frac{eam_{ph}}{m^2} \right)^2 = 10^{-6},
\label{eq:av assump5}
\end{equation}
\begin{equation}
\frac{\mathcal{F}}{\chi^2} = \left(  \frac{mm_{ph}}{k \cdot p} \right)^2  = \left( \frac{m}{p_0}  \right)^2
\label{eq:av assump6}
\end{equation}
and
\begin{equation}
\Omega^2 - 1 = \left( \frac{eam_{ph}}{k \cdot p } \right)^2 = 10^6 \left( \frac{m}{p_0 } \right)^2 .
\label{eq:av assump7}
\end{equation}
One can see that the difference between (\ref{eq:av assump5},\ref{eq:av assump6}) and (\ref{eq:av assump7}) is of several order of magnitudes. As a result, according to (\ref{eq:av assump6}, \ref{eq:av assump7}) the Volkov approximation breaks only if $p_0 \approx m$ while our new condition (\ref{eq:av assump7}) states that the Volkov solution is not valid in the range $m<p_0<ea$, which is extremely relevant for cascades formation.

Let us examine $\delta, \Omega$ as a function of the asymptotic momentum for a given $m_{ph}$. In terms of the physical quantites, $\delta$ (defined in (\ref{eq:av cond1})) takes the form
\begin{equation}
\delta= \frac{2eam_{ph}^2 |p \cdot \epsilon|}{(k \cdot p)^2+ (eam_{ph})^2}.
\label{eq:av delta2}
\end{equation}
In the wave framework,
\begin{equation}
|\epsilon \cdot p| = \sqrt{\frac{p_1^2+p_2^2}{2}}=\frac{p_{\bot}}{\sqrt{2}}  .
\end{equation}
Finally, we have
\begin{equation}
\delta = \frac{\sqrt{2} ea p_{\bot}}{p_0^2+ (ea)^2}, \quad \Omega = \sqrt{1 +  \left( \frac{ea}{p_0 } \right)^2  }  .
\label{eq:av delta3}
\end{equation}
There are several regimes, according to the values of $p_0,ea$ and $p_{\bot}$. \\
A. If $p_0 \ll ea$ we obtain $\delta \ll 1$ and $\Omega \gg 1$. The meaning is that our solution is valid and its deviation from Volkov is significant. For the laser parameters mentioned earlier, $ea/m = 10^3, m_{ph} / m = 10^{-6}$, the value of $\Omega$ lies in the range $1-10^3$.
 \\
B. In the opposite case $p_0 \gg ea$, it follows that $\delta \ll 1$ and $\Omega \approx 1$. Namely, our solution is valid and recovers the Volkov wavefunction. The first correction to the Volkov solution, i.e. the WKB approximation, corresponds to the expansion of $\Omega$ in powers of $m_{ph}$. Notice that in the above derivation, as opposed to the WKB approach, $m_{ph}$ is not assumed to be small and may get any value as long as $\delta \ll 1$.\\
C. If the particle energy is of the same order of magnitude as the field amplitude ($p_0 \approx ea$) there are two possibilities, depending on the value of $p_{\bot}$. On the one hand, $p_{\bot} \ll p_0$ leads to $\delta \ll 1$ and $\Omega \approx \sqrt{2}$, meaning that our solution is valid. On the other hand, $p_{\bot} \approx p_0$, we have $\delta \approx 1$. Consequently, the effective equation (\ref{eq:av effective}) does not represent the original Mathieu equation, giving rise to second order behavior such as bands structure formation. This regime is extremely interesting and will be addressed in a seperate publication.

\section{Numerical Results}
In the following, our novel analytical solution is compared with the Volkov wavefunction and with the exact solution. The goal is twofold - illustrate its deviation from Volkov as well as investigate its accuracy for varying values of the small parameter $\delta$. As described in the previous sections, all three solutions may be cast in the general form (\ref{eq:av sol_final4}) and are therefore characterized by $\nu$ and $c_{2n}$. These quantities are given by (\ref{eq:av n_nu}, \ref{eq:av coeff1}) for our new solution and by (\ref{eq:av n_nu_V}, \ref{eq:av coeff2}) For Volkov. 
As to the exact wavefunction case, they are obtained numerically according to (\ref{eq:av Vn}, \ref{eq:av recursion2} - \ref{eq:av cond2}, \ref{eq:av nu_def}). Notice that adding to $\nu$ any integer $j$ leaves the solution (\ref{eq:av sol_final4}) unchanged besides a shift in the distribution $c_{2n} \rightarrow c_{2n+2j}$. It allows the numerical algorithm, when searching for $\nu_e$, to be restricted to $\nu_e = \bar{\nu}_e$  where $\bar{\nu}_e$ lies the range $0<\bar{\nu}_e<1$. Afterwards, it may be shifted by an integer, $\nu_e = \bar{\nu}_e + j$, so as to make the distribution centered around $n=0$. 
The normalization of the various solutions was determined according to the condition 
\begin{equation}
\int{d^3x} \left[ \Phi_p \partial_0 \Phi_p^* - \Phi_p^* \partial_0 \Phi_p \right] = 1.
\end{equation}

For the sake of demonstration only, the plots appearing below were calculated with different laser parameters than these of ELI mentioned in the previous section ($ea/m = 10^3, m_{ph} / m = 10^{-6}$). The reason lies in fact that these parameters result in an enormous number of harmonics (as one obtains by substituting them into (\ref{eq:av n_nu}))
\begin{equation}
n^* =10^9 \frac{ p_{\bot}}{\Omega p_0}, \quad 1< \Omega <1000
\end{equation}
making the wavefunction graphically difficult for inspection. As a result, the laser parameters were chosen to be $\xi=20, m_{ph}=m/10$, yielding the range
\begin{equation}
n^* =2000 \frac{ p_{\bot}}{\Omega p_0}, \quad 1< \Omega <20.
\end{equation}
Nevertheless, the physical effect we wish to demonstrate remains the same.

We start with calculation parameters corresponding to $\delta \ll 1$, i.e. small transverse momentum $p_{\bot} \ll p_0,ea$ (see Eq. (\ref{eq:av delta3})). We have taken $\textbf{p}=(m/5,m/5,0)$, corresponding to $\Omega \approx 19.3$ and $\delta = 0.02$. 
The spectral shape of the Volkov and the new wavefunctions is shown in Fig. 1. The exact wavefunction was calculated as well, but was not plotted in the figure as its deviation from our new analytical solution are extremely negligble. 
The deviation from the Volkov wavefunction is overwhelming - the width of the new solution is $\approx 20$ times smaller than that of Volkov, in agreement with the value of $\Omega$.
It should be stressed that each distribution is centered around a different $\nu$. For Volkov we have $\nu_V = 1924.5$, while our solution corresponds to $\nu = 	 189.9 $. 

Fig. 2 depicts the exact, analytical and Volkov wavefunctions for $\textbf{p}=(5m,5m,0)$, corresponding to $\delta = 0.44, \Omega = 2.97$. Several interesting points stem from the comparison. First, the difference between the spectral width of the Volkov and the analytical solutions decreases with respect to the previous case. Second, the analytical solution is fairly close to the exact one, even though $\delta$ is not negligible. Third, the spectral shape of the exact solution is deformed, implying that the symmetry between photon emission and absorption no longer exists. 
The distributions are centered arround $\nu_V = 	 280.06 $, $\nu = 	 140.3 $ for the Volkov wavefunction and the new solution respectively. For the exact wavefunction, however, the non-symmetric distribution implies that the definition of $\nu $ as the center of the distribution loose its meaning. 
Hence, we have chosen $\nu_e$ to be as close as possible to the analytical solution, in order to ease the comparison. The numerical calculation yielded $\nu_e = 0.97$ and was shifted, according to the above argument, to $\nu_e = 	 140.97$.
\begin{figure}[H]
  \begin{center}
   	   \includegraphics[width=0.5\textwidth]{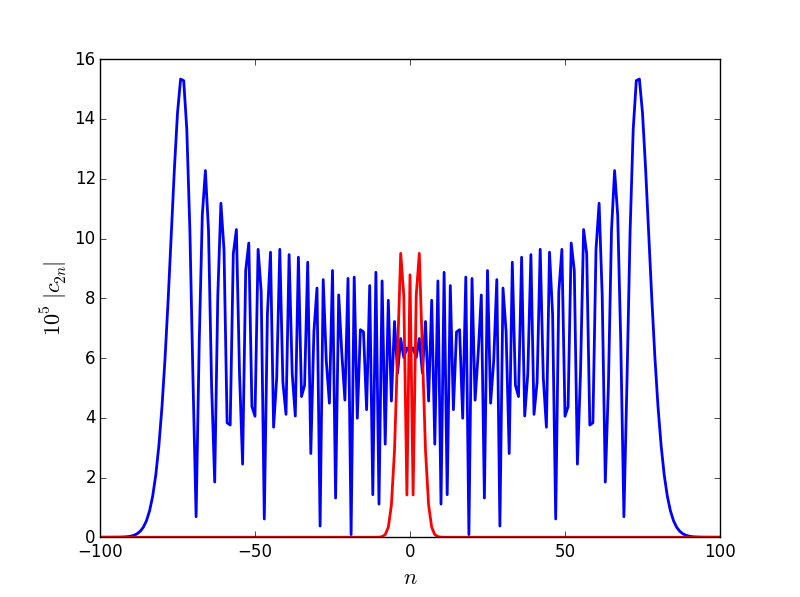}
      \caption{(color online). The wavefucntion spectral coefficients of the Volkov (blue curve) and the analytical (red curve) solutions respectively, for $\xi=20, m_{ph}=m/10, \textbf{p} = (0.2m,0.2m,0)$, corresponding to $\delta = 0.02$.}
  \end{center}
\end{figure}
\begin{figure}[H]
  \begin{center}
   	   \includegraphics[width=0.5\textwidth]{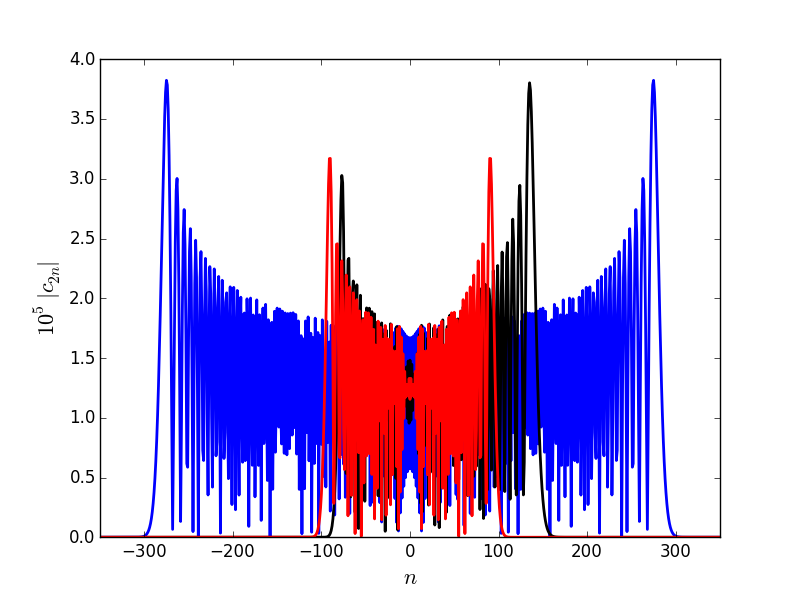}
      \caption{(color online). The wavefucntion spectral coefficients of the Volkov (blue curve), the analytical (red curve) and the exact (black curve) solutions respectively, for $\xi=20, m_{ph}=m/10, \textbf{p} = (5m,5m,0)$, corresponding to $\delta = 0.44$.}
  \end{center}
\end{figure}

\section{Conclusion}
The equation of motion of a particle in a rotating electric field, taking the form of the Mathieu equation, was analyzed. A novel approximated solution was found, adequate if the asymptotic energy of the particle (in the wave frame) is much smaller or much higher than the field amplitude.
For the first case, $p_0 \ll ea$, the spectral width of the analytical solution (\ref{eq:av sol_final2}) was shown to be $\Omega$ times smaller than the width of a Volkov wavefucntion with the same $p_{\mu}$. 
For the second case, $p_0 \gg ea$, corresponding to an energetic electron beam colliding with a laser, our solution recovers the Volkov wavefunction. The differences between Volkov, our new solution and the exact solution in the intermediate regime $p_0 \approx ea$ were explored numerically.

As described in the introduction, the emission rates embedded in standard QED cascade calculations rely upon the Volkov wavefunction and depend on the variables $\xi, \chi$. The underlying assumption is that the Volkov solution is applicable as long as (\ref{eq:av assump}) holds.
However, according to the above analysis, deviations from the Volkov solution occur unless $p_0 \gg ea$, in contradiction with (\ref{eq:av assump}).
Consequently, we argue that the emission processes are no longer described by the well known expressions obtained with the Volkov wavefunction.
Moreover, they depend upon another parameter, taking into account the value of $m_{ph}$. The modifed rates may be obtained with our novel solution presented above. The apparent similarity between the mathematical structure of our solution and the Volkov wavefunction implies that the mathematical techniques exploited to derive the Volkov rates may be of use in our case as well. 
The predicted cross section of such a calculation is supposed to be of quantum nature and to deviate from Volkov provided that $\chi \approx 1$ and $p_0 < ea$.
For optical lasers ($m_{ph}/m \approx 10^{-6}$) it follows from (\ref{eq:av chi2}) that $ea/m \approx 10^3$ (corresponding to $I \approx 10^{24}W/cm^2$) is required. 
Consequently, our solution may be put to test with the next generation laser systems.





\end{document}